# Topologically Nontrivial Interband Plasmons in Type-II Weyl Semimetal $MoTe_2$

Xun Jia,[1,2,*] Maoyuan Wang,[3,*] Dayu Yan,[1,2,*] Siwei Xue,[1,2] Shuyuan Zhang,[1,2] Jianhui Zhou,[4,†] Youguo Shi,[1,5] Xuetao Zhu,[1,2,5,‡] Yugui Yao,[3] Jiandong Guo[1,2,5,6,§]

[1]*Beijing National Laboratory for Condensed Matter Physics and Institute of Physics, Chinese Academy of Sciences, Beijing 100190, China*

[2]*School of Physical Sciences, University of Chinese Academy of Sciences, Beijing 100049, China*

[3]*Key Laboratory of advanced optoelectronic quantum architecture and measurement (MOE), Beijing Key Laboratory of Nanophotonics and Ultrafine Optoelectronic Systems, School of Physics, Beijing Institute of Technology, Beijing 100081, China*

[4]*Anhui Province Key Laboratory of Condensed Matter Physics at Extreme Conditions, High Magnetic Field Laboratory, Chinese Academy of Sciences, Hefei 230031, China*

[5]*Songshan Lake Materials Laboratory, Dongguan, Guangdong 523808, China*

[6]*Collaborative Innovation Center of Quantum Matter, Beijing 100871, China*

**Abstract**:

**In many realistic topological materials, more than one kind of fermions contribute to the electronic bands crossing the Fermi level, leading to various novel phenomena. Here, using momentum-resolved inelastic electron scattering, we investigate the plasmons and their evolution across the phase transition in a type-II Weyl Semimetal $MoTe_2$, in which both Weyl fermions and trivial nonrelativistic fermions contribute to the Fermi surface in the $T_d$ phase. One plasmon mode in the 1T' phase at high temperature and two plasmon modes in the topological $T_d$ phase at low temperature are observed. Combining with first-principles calculations, we show that all the plasmon modes are dominated by the interband correlations between the inverted bands of $MoTe_2$. Especially in the $T_d$ phase, since the electronic bands split due to inversion symmetry breaking and spin-orbit coupling,**

**the plasmon modes manifest the interband correlation between the topological Weyl fermions and the trivial nonrelativistic electrons. Our work emphasizes the significance of the interplay between different kinds of carriers in plasmons of topological materials.**

# I. Introduction

Topological materials including topological insulators [1-4] and topological semimetals [1-5] have become an extremely attractive branch in condensed matter physics. The nontrivial topology of Bloch bands could lead to a variety of novel physical phenomena such as the topologically protected edge states and the related unique transport and optical behaviors [6], leading to potential technological applications in spintronics and quantum computations. The collective excitation of electrons due to their long-range Coulomb interaction, *i.e.*, plasmon, is a fundamental concept in physics, and has attracted increasing attentions in the field of topological materials recently. Various novel plasmons have been predicted in topological insulators [7-10] and in topological semimetals [11-23]. For example, the surface states of three-dimensional topological insulators with the spin-momentum locking was predicted to support the spin plasmon, linking spintronics and plasmonics [7]. The Weyl semimetals without time-reversal symmetry were predicted to host the nonreciprocal magnetoplasmon, which is the signature of the chiral anomaly [15]. Importantly, a series of significant experimental progresses on plasmon excitations have been made in topological insulators [24-28] as well as in topological semimetals [29-32], which could greatly facilitate the promising applications of these topological materials in plasmonics [33].

Most of these fascinating plasmons in topological materials can be well described within a single band picture with only one kind of carriers involved. This is similar to the cases in normal metals, in which the intraband correlation dominates [34]. Consequently, the origin of these plasmons can be directly understood in terms of the simple Fermi surfaces. However, in many topological materials, such as NbP and (W, Mo)$Te_2$, the topologically nontrivial bands and the trivial bands coexist around the Fermi level, and thus more than one kind of fermions contribute to the Fermi surface [35-44]. As the result, the plasmon modes in these systems may be contributed from multiple kinds of Fermions. Especially, when the topologically nontrivial and trivial bands highly mix with each other, their couplings could significantly affect the properties of the plasmons, which is beyond the scope of the single band picture. Therefore, the exotic plasmon properties based on the assumption of pure topological bands may not be applicable in real topological materials.

In this paper, using high-resolution electron energy loss spectroscopy (HREELS) [45], we studied the plasmons in $MoTe_2$, which exhibits the structural phase transition from the high temperature monoclinic 1T' phase to the orthorhombic $T_d$ phase with the critical temperature of ~250 K. We observe one plasmon mode in the 1T' phase and two plasmon modes in the $T_d$ phase. The energies of these plasmon modes are almost dispersionless and exhibit distinct nonlinear temperature dependences. Combining with first-principles calculations, we reveal that unlike the conventional plasmons dominated by the intraband correlations, all these plasmon modes in $MoTe_2$ mainly originate from the interband correlations. Especially in the $T_d$ phase, the two plasmon modes are dominated by the correlations between the nontrivial Weyl Fermion bands and the trivial nonrelativistic electronic bands. Our work clearly demonstrates that the interplay of nonrelativistic fermions and Weyl fermions plays a crucial role in plasmons of realistic topological materials.

# II. Materials and methods

## A. Crystal Preparation

Single crystal of 1T'-$MoTe_2$ were grown by using Te as flux. Starting materials Mo (Column, 99.9999%) and Te (Lump, 99.9999%) were mixed in an Ar-filled glove box at a molar ratio of Mo : Te = 1 : 20. The mixture was placed in an alumina crucible, which was then sealed in an evacuated quartz tube. The tube was heated to 1100 °C over 20 hours and dwelt for 10 hours. Then, the tube was slowly cooled down to 950 °C at a rate of 1 °C/h followed by separating the crystals from the Te flux by centrifuging. Shiny crystals with the size of 1×5 $mm^2$ were obtained on the bottom of the crucible.

The good crystal quality is characterized *ex situ* by X-ray diffraction (XRD) and the surface of the cleaved sample is checked by X-ray photoelectron spectroscopy (XPS), which can exclude the possible surface contamination or oxidation of them. The details of the sample characterizations are shown in Appendix A.

## B. HREELS Measurement

As a surface sensitive technique, HREELS is an ideal candidate to explore the low-energy collective excitations of $MoTe_2$. Compared with conventional HREELS, our recently developed two-dimensional (2D)-HREELS can directly obtain a 2D energy-momentum mapping in a very large momentum scale without rotating sample, monochromator, or analyzer [46].

The energy and momentum of the collective excitations (either plasmon or phonon) are obtained using the conservation of energy and momentum for incident and scattered electrons. As given by $\hbar q_{\parallel} = \hbar(k_i \sin\alpha_i - k_s \sin\alpha_s)$ (where $\alpha_i$ and $\alpha_s$ are the incident and scattering angles, respectively), the parallel momentum $q_{\parallel}$ depends on incident energy $E_i$, energy loss $E_{loss} = E_s - E_i$, $\alpha_i$ and $\alpha_s$ according to

$$q_{\|} = \frac{\sqrt{2mE_i}}{\hbar}(\sin\alpha_i - \sqrt{1-\frac{E_{loss}}{E_i}}\sin\alpha_s) \approx \frac{\sqrt{2mE_i}}{\hbar}(\sin\alpha_i - \sin\alpha_s) \quad (1)$$

In this study, all the HREELS measurements were performed in situ within ~10 hours after fresh cleavage in ultra-high vacuum (~$1\times10^{-10}$ Torr). We obtained the information around the first Brillouin zone center Γ with the incident energy ranging from 15 eV to 110 eV at room temperature. And the temperature-dependent measurements were obtained with the incident energy of 110 eV only.

### C. Details of First-principles Calculations

The first-principles calculations are performed by using Vienna ab initio simulation package (VASP) [47] based on the density function theory with Perdew-Burke-Ernzerhof (PBE) parameterization of generalized gradient approximation (GGA) [48,49]. The energy cutoff of the plane wave basis is set as 300 eV, and the Brillouin zone of $MoTe_2$ is sampled by 18 × 10 × 4 k-mesh. The ionic positions are fully optimized with the vdW correction until the force on each atom was less than 0.01 eV/Å while the lattice parameters are fixed as the experimental value. The dynamical dielectric function are calculated based on the Hamiltonian of maximally localized Wannier functions (MLWF) [50] with s, p, d orbitals of Mo atoms and s, p orbitals of Te atoms.

### D. Calculations of Dynamical Dielectric Function

Within the random phase approximation (RPA), the dynamical dielectric function $\varepsilon^{\mathrm{RPA}}(q,\omega)$ can be calculated from the formula:

$$\begin{aligned} \varepsilon(q,\omega) &= 1 - \frac{4\pi e^2}{\kappa q^2}\chi_0 \\ &= 1 - \frac{4\pi e^2}{\kappa q^2 V N_k}\sum_{mn,k}\frac{f_{mk}-f_{nk}}{E_{mk}+\omega-E_{nk}+i\eta}\langle mk|nk+q\rangle\langle nk+q|mk\rangle \quad (2) \\ &= 1 - \frac{4\pi e^2}{\kappa q^2 V N_k}\sum_{mn,k}\frac{f_{mk}-f_{nk+q}}{E_{mk}+\omega-E_{nk+q}+i\eta}\langle mk|nk+q\rangle\langle nk+q|mk\rangle \end{aligned}$$

where the effective background dielectric constant $\kappa$ is calculated from VASP, the values of which are 218.2 for the $T_d$ phase and 190.3 for the 1T' phase along the $\Gamma - X$ direction ($\kappa_{xx}$). The formula of $\varepsilon^{\mathrm{RPA}}(q,\omega)$ can be written as the long wavelength limit of $q$ = 0:

$$\varepsilon(0,\omega) = 1 + \frac{4\pi e^2}{\kappa V N_k}\left[\sum_{m,k}\frac{\frac{\partial f_{mk}}{\partial E_{mk}}\left(\frac{\partial E_{mk}}{\partial k}\right)^2}{(\omega+i\eta_1)^2} - \sum_{m,k}\sum_{n\neq m}\frac{f_{mk}-f_{nk}}{E_{mk}-E_{nk}+\omega+i\eta_2}\frac{\langle mk|\frac{\partial H}{\partial k}|nk\rangle\langle nk|\frac{\partial H}{\partial k}|mk\rangle}{(E_{nk}-E_{mk})^2}\right] \quad (3)$$

where the first term is the intraband term with a broadening parameter $\eta_1$ and the second term is the interband term with a broadening parameter $\eta_2$.

# III. Results and Discussions

### A. Crystal Structures of $MoTe_2$

$MoTe_2$ is a layered van der Waals material with two different structures at 300 K: hexagonal (2H) phase or monoclinic (1T') phase, due to different growth conditions. In this study we focus on the 1T' phase (a trivial metal/semimetal), which exhibits a structural phase transition to the $T_d$ phase at ~ 250 K [51]. Although these two phases share the same in-plane structure, the $T_d$ phase with a vertical (90°) stacking belongs to the non-centrosymmetric space group $Pmn2_1$ while the 1T' phase with a distorted stacking belongs to the centrosymmetric space group $P2_1/m$ [see Figs. 1(a) and 1(c)]. Only the low temperature $T_d$ phase was predicted to support type-II Weyl fermions, while the centrosymmetric 1T' phase is topologically trivial. Angle-resolved photoemission spectroscopy together with the first-principles calculations shows that Fermi surfaces in the $T_d$ phase $MoTe_2$ consist of both topologically trivial and nontrivial bands [39-42,44]. Fig. 1(e) shows the Brillouin zone (BZ) of the $T_d$ phase, and the location of the Weyl points are shown in Fig. 1(f) and a corresponding enlarged zone in Fig. 1(g). Previous theoretical and experimental works suggest four pairs of Weyl points in the $k_z$ = 0 plane of the BZ [35,36].

The $MoTe_2$ single crystals with the 1T' phase at ~294 K were cleaved *in situ*, and the obtained (001) surface exhibits excellent (1 × 1) low energy electron diffraction (LEED) patterns of sharp spots with a very low background, as shown in Fig. 1(b). At 44 K, the $T_d$ phase surface shows almost unchanged (1×1) LEED patterns [Fig. 1(d)] as the 1T' phase, since the two phases share the same in-plane structure. The angle-resolved inelastic electron scattering was performed by the HREELS with the capability of 2D energy-momentum mapping [46]. The results in both the 1T' and $T_d$ phases at different temperatures were obtained.

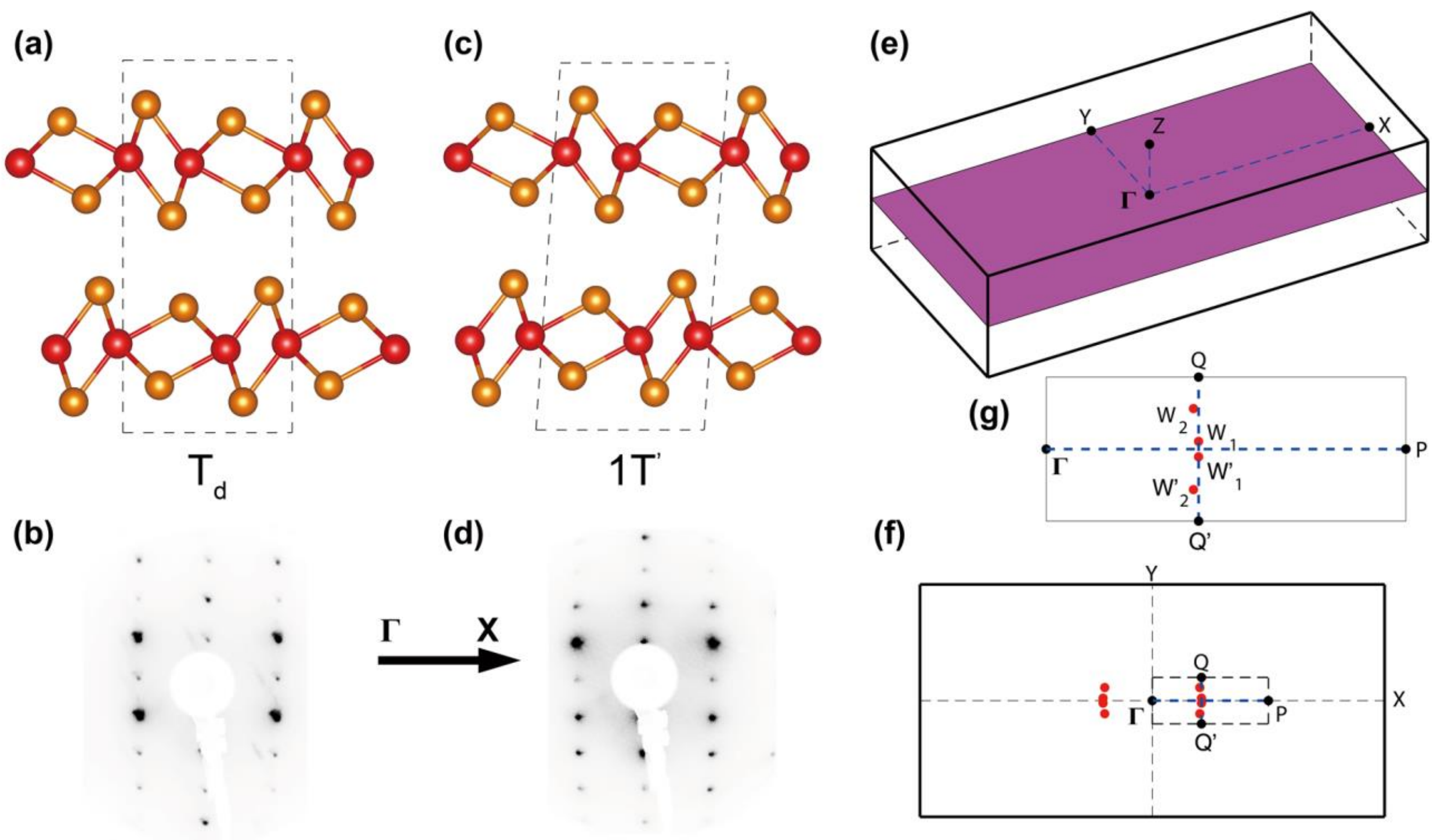

Fig. 1. Schematic illustration of the $T_d$ phase and 1T' phase of $MoTe_2$. (a) and (c) crystal structures of the $T_d$ phase and 1T' phase, respectively. (b) and (d) LEED patterns of the $T_d$ phase and 1T' phase, respectively. (e) Brillouin zone of the $T_d$ phase, where pink part is the $k_z$ = 0 plane, which is also plotted in (f). The red dots in (f) are eight Weyl points in the $T_d$ phase, and (g) shows four of them in an enlarged zone.

### B. Plasmons from HREELS Measurements

Figures 2(a) and 2(b) display the 2D energy-momentum mappings obtained from the HREELS measurements along the $\Gamma - X$ direction with the incident beam energy of 110 eV for the $T_d$ phase at 44 K and for the 1T' phase at 294 K, respectively. In the $T_d$ phase, the energy distribution curve (EDC) integrated over momentum [red curve in Fig. 2(a)] shows two distinct energy-loss peaks around 90 and 170 meV, which are labeled as α and β, respectively. In contract, the integrated EDC for the 1T' phase only shows one energy-loss peak around 220 meV, which is labeled as γ. The energies of all these peaks are much higher than the highest phonon energy (around 35 meV) in $MoTe_2$ [52], indicating that they are not phonons. Through XPS measurement to test the 3d binding energy of Te and Mo, and the comparison of the HREELS spectra between clean and exposed surfaces, we show that these loss peaks are not vibrations of possibly adsorbed molecules (*e.g.*, $H_2O$) or the Te-O bond due to possible surface oxidation. Instead, they are most likely to be plasmons.

Then the dispersions of these modes are checked from $q$-dependent EDCs extracted from the 2D mapping, as shown in Figs. 2(c) and 2(d). Due to the semimetallic nature of $MoTe_2$, there exists a strong Drude background in the energy loss EDCs. We employ a background subtraction method based on a polynomial fitting of the baseline, which has been used in graphene [53] and graphite [54], to extract the information of the exact energy loss peaks, including energy, full width at half maximum (FWHM), and intensity (see details in Appendix B). The obtained experimental energy-momentum points of these modes are plotted in Figs. 2(e-h) (solid dots). All of the three modes are almost dispersionless, and their intensities show quick damping with the increasing momentum $q$, becoming invisible when $q > 0.05$ Å$^{-1}$. In addition, their energies are nonzero at $q = 0$, evidencing that they are plasmons originating from bulk bands. In the $T_d$ phase, the average ratio of energy between α and β is $\frac{\omega_\alpha}{\omega_\beta} \sim 0.56 \pm 0.04$, significantly deviating from the conventional ratio between the energies of bulk plasmon (bp) and its corresponding surface plasmon (sp) from the same electronic band: $\frac{\omega_{\mathrm{sp}}}{\omega_{\mathrm{bp}}} = \frac{1}{\sqrt{2}} \sim 0.71$ for the ideal case, and varies from 0.60±0.35 (Lithium) to 0.73±0.11 (Potassium) for typical real metals [55]. Additionally, $\frac{\omega_\alpha}{\omega_\beta}$ is almost independent on the incident angle of the electron beam and the incident energy (details shown in Appendix C), suggesting that β and α cannot be the conventional bulk plasmon and its corresponding surface plasmon.

To further clarify the nature of these modes, we performed first-principles calculations to obtain the electronic bands as well as the density-density correlation functions in the two phases of $MoTe_2$. Then the dynamical dielectric functions $\varepsilon^{\mathrm{RPA}}(q,\omega)$ were calculated within the random phase approximation (RPA). The calculated plasmon modes appear as sharp peaks in the energy loss function $\varepsilon_{\mathrm{loss}} = -\mathrm{Im}(\frac{1}{\varepsilon^{\mathrm{RPA}}})$, which are directly related to the experimental excitation spectra probed by HREELS. In addition to the terms describing the intraband transitions in the regular plasmon study, the calculations also include the terms describing the interband transitions which may dominate in the plasmons with inverted energy bands [9]. We set the Fermi energy at -50 meV, and use two different broadening parameters for the intraband ($\eta_1$ = 150 meV) term and the interband ($\eta_2$ = 20 meV) term, respectively, to fit the experimental results. It turns out that the interband transitions dominate the loss function intensities, while the intraband transitions only act as negligible backgrounds. We will focus on the interband contributions in the following. The intensity mappings of the calculated electron energy loss functions for the $T_d$ phase and the 1T' phase are shown in Figs. 2(e) and 2(f), and the corresponding loss functions at different $q$ are plotted in Figs. 2(g) and 2(h), respectively. The experimental energy-momentum points of the three modes are superimposed on the calculated results. The theoretical simulation in the small $q$ range shows a good agreement with the measured peak positions. The main discrepancy is that the calculated results damp slower and disperse to a larger momentum range where the loss peaks are invisible in the measurements. This is largely due to the simplification of the theoretical model. In the simulations, all the plasmon modes are simplified as interband correlations, with the intraband correlations neglected, which might also contribute to the damping rate. Other scattering processes, such as electron-phonon scattering, electron-defect scattering and electron-surface scattering, are also ignored in the simulations, resulting in the discrepancy of the momentum of electron-hole excitations between the simulation and experimental observations. Nevertheless, the theoretical calculations can well reproduce the energies of the observed plasmon modes as well as their connections to the phase transition. Details of the calculation methods and analyses can be found in Appendix D.

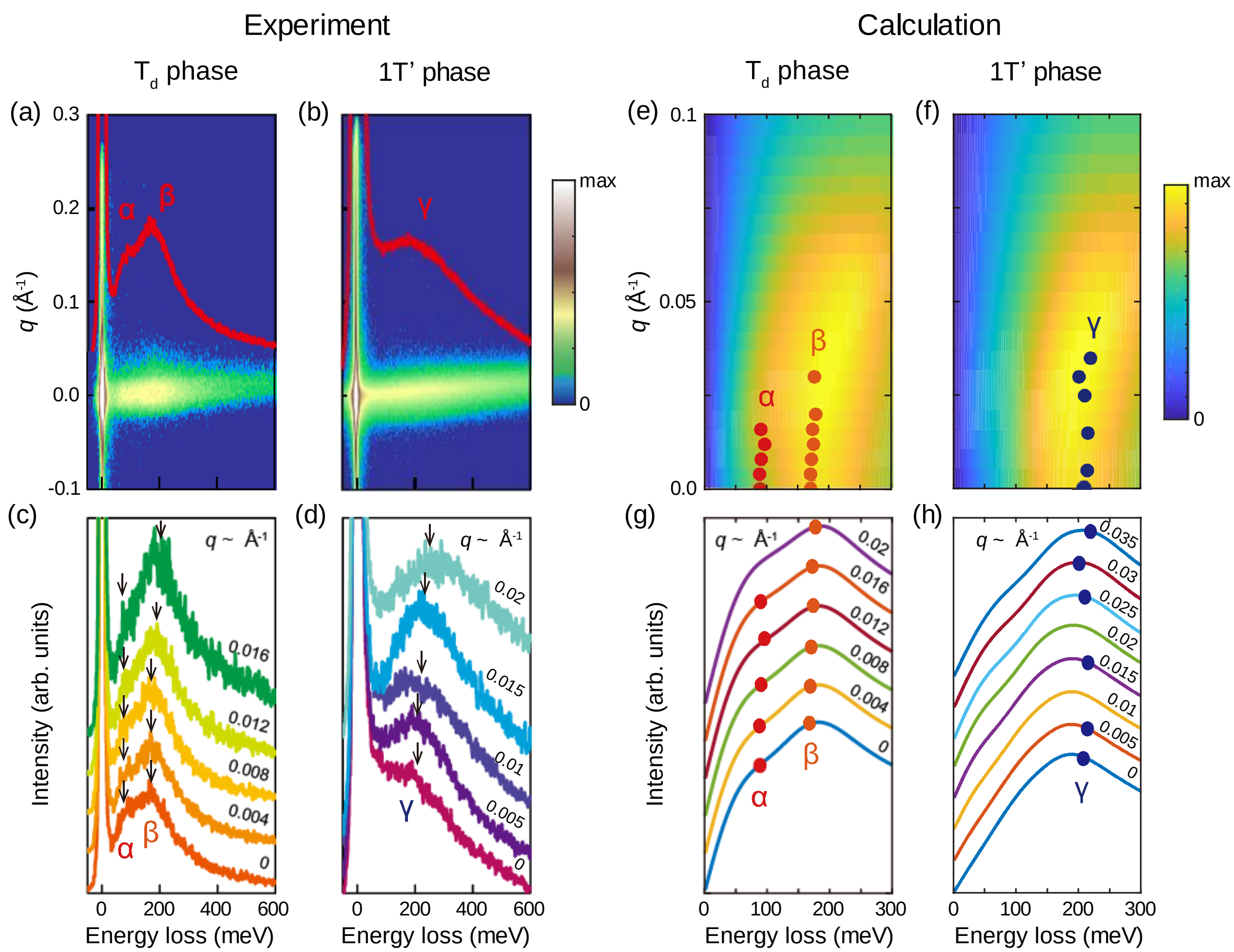

Fig. 2. Experimental and calculated plasmon modes of $MoTe_2$ in the $T_d$ phase at 44 K and 1T' phase at 294 K. (a) and (b) Experimental 2D HREELS energy-momentum mapping of the $T_d$ phase and 1T' phase, respectively, both along the $\Gamma - X$ direction with incident energy of 110 eV. The red curves superimposed on the 2D mapping are EDCs integrated over the momentum. (c) and (d) EDCs at different momentum values extracted from (a) and (b), respectively, with the arrows indicating the peak positions. (e) and (f) 2D momentum-resolved loss function mapping calculated by DFT in the $T_d$ phase and 1T' phase, respectively. The red and orange dots in (e) and the blue dots in (f) represent the experimental data extracted from the EDCs. (g) and (h) Comparison of the calculated loss functions with the measured peak positions.

## C. Temperature-dependence of Plasmons

To gain more insights into the behaviors of plasmons accompanied with the structural phase transition, we performed temperature-dependent HREELS measurements from 294 K to 44 K. Figure 3(a) shows the stacked EDCs at Γ point at several temperatures (a complete set of temperature-dependent measurements is provided in Appendix C). With the temperature decreasing, the γ mode gradually evolves into the β mode, while the α mode gradually appears. The energies and intensities of these three modes as a function of temperature are extracted by using the fitting method mentioned above, with the results plotted in Figs. 3(b) and (c). When the temperature is above ~260 K, the data can be only well fitted with one peak. And when the temperature is below ~200 K, the data can be only well fitted with two peaks. It should be noted that, in the temperature range from 200 K to 260 K where the structural phase transition temperature $T_C$ locates, the data can be fitted equally well either with one peak or with two peaks. This temperature range was marked by a gray rectangle in Figs. 3(b) and 3(c) to show the uncertainty of the data fitting. The present HREELS plasmon measurements cannot provide accurate information near $T_C$, possibly because of the intermediate new phase with frustrated disorders [56] in the structure associated with the phase transition. But it is pretty clear that only the γ mode exists in the high-temperature 1T' phase, while there are indeed two modes (α and β) existing in the low-temperature $T_d$ phase.

Figure 3(b) shows that the energies of the observed plasmon modes exhibit different temperature-dependent behaviors. In the $T_d$ phase below $T_C$, the energy of the β mode shows a very weak temperature-dependence, while the α mode shows a very strong nonmonotonic temperature-dependence. In the 1T' phase above $T_C$, the energy of the γ mode shows very slight increase with increasing temperature. It has been theoretically predicted that the energy of the intrinsic plasmon in type-I Weyl/Dirac semimetals has a non-linear temperature-dependence when only a single Weyl/Dirac cone contributes to the Fermi surface [13,14], similar to the temperature dependence of the α mode in our experiment. However, since the electronic structure of $MoTe_2$ is more complicated than the single Weyl/Dirac cone model, and all of the three plasmon modes mainly originate from the interband correlations. As a result, the different temperature dependences of energies of these modes should be hard to be faithfully captured by the theoretical calculations based on the simple ideal model.

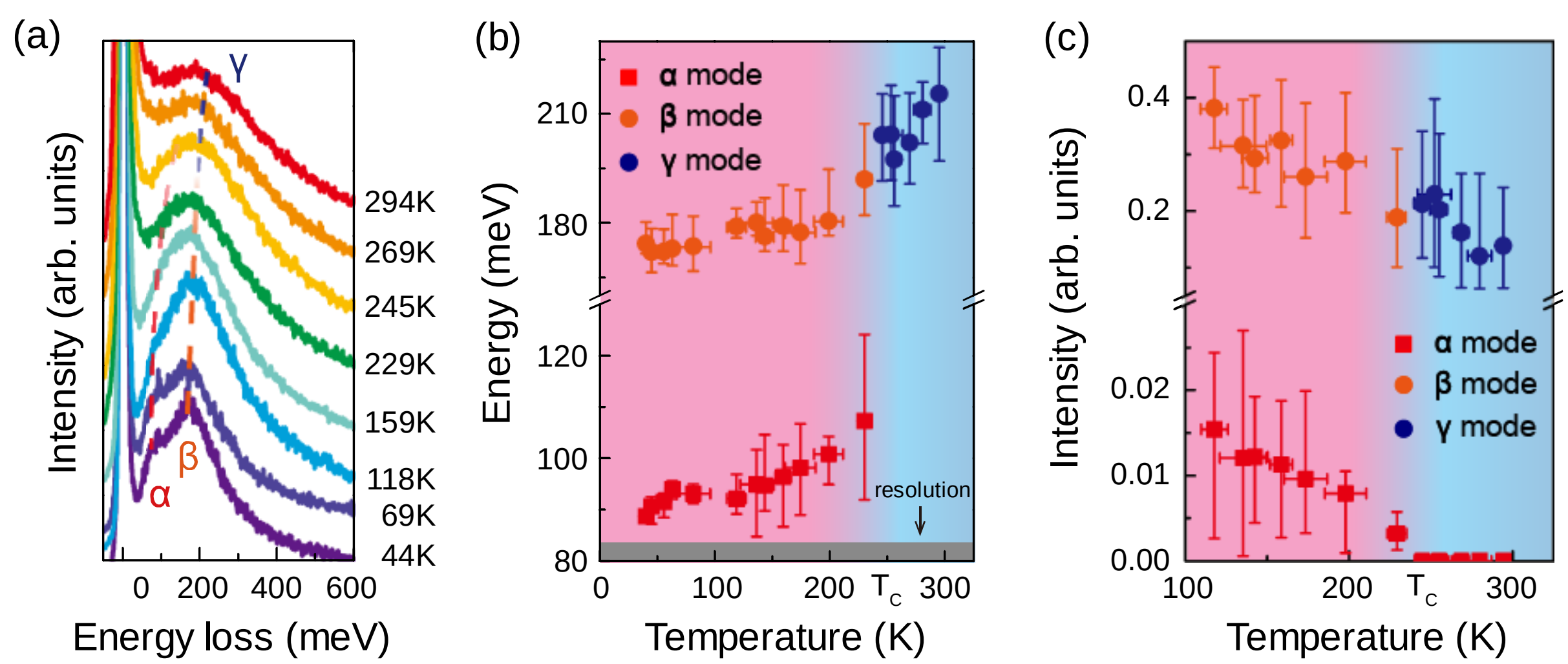

Fig. 3. Temperature dependence of the HREELS results of $MoTe_2$. (a) EDCs at Γ point with temperatures from 44 K to 294 K. The positions of α, β and γ modes are labeled by gradually varied dashed red, orange, and blue lines, respectively, as guides to the eye. (b) Temperature dependence of the energies of α, β and γ modes at Γ point. The gray band at the bottom represents the energy resolution of our facility in this study. (c) Temperature dependence of the normalized intensities of α, β and γ modes at Γ point. The error bars in (b) and (c) are generated by different baselines used in the background subtraction during the data fitting.

### D. Origin of the Interband Plasmons

To explore the origin of these plasmon modes, we decompose the contribution of each plasmon mode to the energy loss function [*i.e.*, $\mathrm{Im}\,\varepsilon^{\mathrm{RPA}}(0,\omega)$] (with details described in the Supplemental Materials) and project it onto the energy bands along a selected path in the BZ. Figures. 4(a) - 4(c) show the distributions of $\mathrm{Im}\,\varepsilon^{\mathrm{RPA}}(0,\omega)$ for α, β and γ in the BZ, respectively. The distributions of $\mathrm{Im}\,\varepsilon^{\mathrm{RPA}}(0,\omega)$ for the α and β modes are different, and those for the β and γ modes share some similarity, while they are still distinct due to the difference of the band structures in the two phases. Figures. 4(d) - 4(g) illustrate the band structures, contributions and the corresponding strength of $\mathrm{Im}\,\varepsilon^{\mathrm{RPA}}(0,\omega)$ for the three modes in the $T_d$ phase and 1T' phase along the Γ - P direction and Q' - Q direction, respectively. The relevant bands are labeled as *a*, *b*, *c* in the 1T' phase and $a_1$, $a_2$, $b_1$, $b_2$, $c_1$, $c_2$ in the $T_d$ phase. As shown in Fig. 4(f), the band *a* and the band *c* can be regarded as two crossing band with a band gap under the influence of the spin-orbit coupling. In this kind of inverted band structure, a plasmon mode γ originating from the interband transitions indeed exists, similar to the case in Ref. [9]. Accompanied with the structural phase transition from the 1T' phase to the $T_d$ phase, the bands *a*, *b*, *c* split into $a_1/a_2$, $b_1/b_2$, and $c_1/c_2$ due to the inversion symmetry breaking, forming eight Weyl points at the crossing points of the bands $a_1/a_2$ and $b_1/b_2$. Accordingly, the γ mode of the 1T' phase splits into two plasmon modes, α and β. The interband correlations between $a_2$ and $c_1$ contribute to the α mode while the correlations between $a_1$ and $c_2$ contribute to the β mode [shown in Fig. 4(d)]. These two plasmon modes in the $T_d$ phase are dominated by the interband correlations between the topologically nontrivial bands and trivial bands.

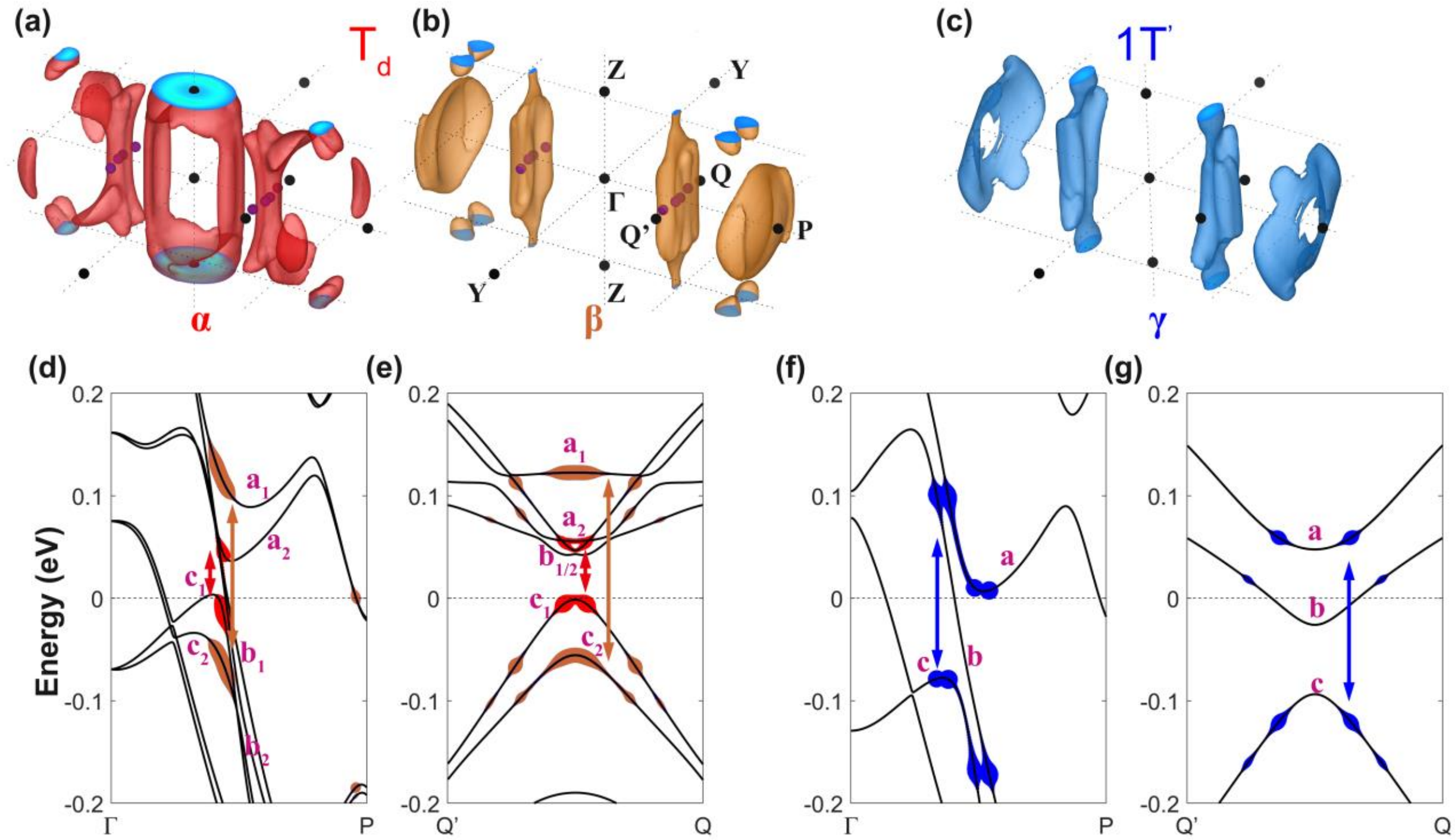

Fig. 4. The physical origin of the plasmon modes α, β and γ. (a), (b) and (c) The distribution of the imaginary part of dielectric function $\mathrm{Im}\,\varepsilon^{\mathrm{RPA}}(0,\omega)$ for α, β and γ in the BZ, respectively. The black dots are some *k*-points in the BZ, while Γ, X, Y, and Z are high symmetry points. And the pink dots are the eight Weyl points in the $k_z = 0$ plane. (d) - (g) band structures of the $T_d$ phase and 1T' phase in two directions, respectively. Pink letters mark different bands, and colored dots are used to mark the contribution strength of the bands for $\mathrm{Im}\,\varepsilon^{\mathrm{RPA}}(0,\omega)$ in the plasmon modes α (red), β (orange) and γ (blue). The arrows illustrate the corresponding interband correlations.

# IV. Conclusions

In conclusion, we have systematically investigated the plasmon modes of $MoTe_2$ at different temperatures. Our HREELS experiments indicate the presence of two plasmons (α and β) in the low temperature $T_d$ phase and only one mode (γ) in the high temperature 1T' phase. Combining with first-principles calculations, we find that all the modes are dominated by interband correlations between the inverted bands of $MoTe_2$. Especially, the modes in the topological $T_d$ phase, dominated by the interband correlations between the topologically nontrivial bands and trivial bands, are attributed to the band splitting due to inversion symmetry breaking. This work reveals the limitation of the single band picture and significantly broadens the understanding of plasmon modes in realistic topological materials, in which band mixing usually exists.

## Acknowledgements

The work was supported by the National Natural Science Foundation of China (Nos. 11874404, and 11634016), the National Key R&D Program of China (Nos. 2016YFA0302400, and 2017YFA0303600). X.Z. was partially supported by the Youth Innovation Promotion Association of Chinese Academy of Sciences (No. 2016008). J.G. was partially supported by BAQIS Research Program (No. Y18G09). Y.S. was partially supported by the Beijing Natural Science Foundation (No. Z180008). J.Z. was supported by the High Magnetic Field Laboratory of Anhui Province. M.W. and Y.Y. were supported by the National Key R&D Program of China (No. 2016YFA0300600), the National Natural Science Foundation of China (No. 11734003), and the Strategic Priority Research Program of Chinese Academy of Sciences (No. XDB30000000).

*These authors contributed equally to this work.

Corresponding authors: †jhzhou@hmfl.ac.cn; ‡xtzhu@iphy.ac.cn; §jdguo@iphy.ac.cn

## Appendix A: Sample Characterization

Figure 5 shows the X-ray diffraction (XRD) pattern measured from the (001) surface of $MoTe_2$ single crystal that we used in our HREELS study. The XRD pattern shows a series of (00l) peaks, indicating good crystalline quality. By comparing with the reported XRD data in the existing studies of $MoTe_2$ topological properties [39,57,58], the crystalline quality in our study is better than most of the reported samples.

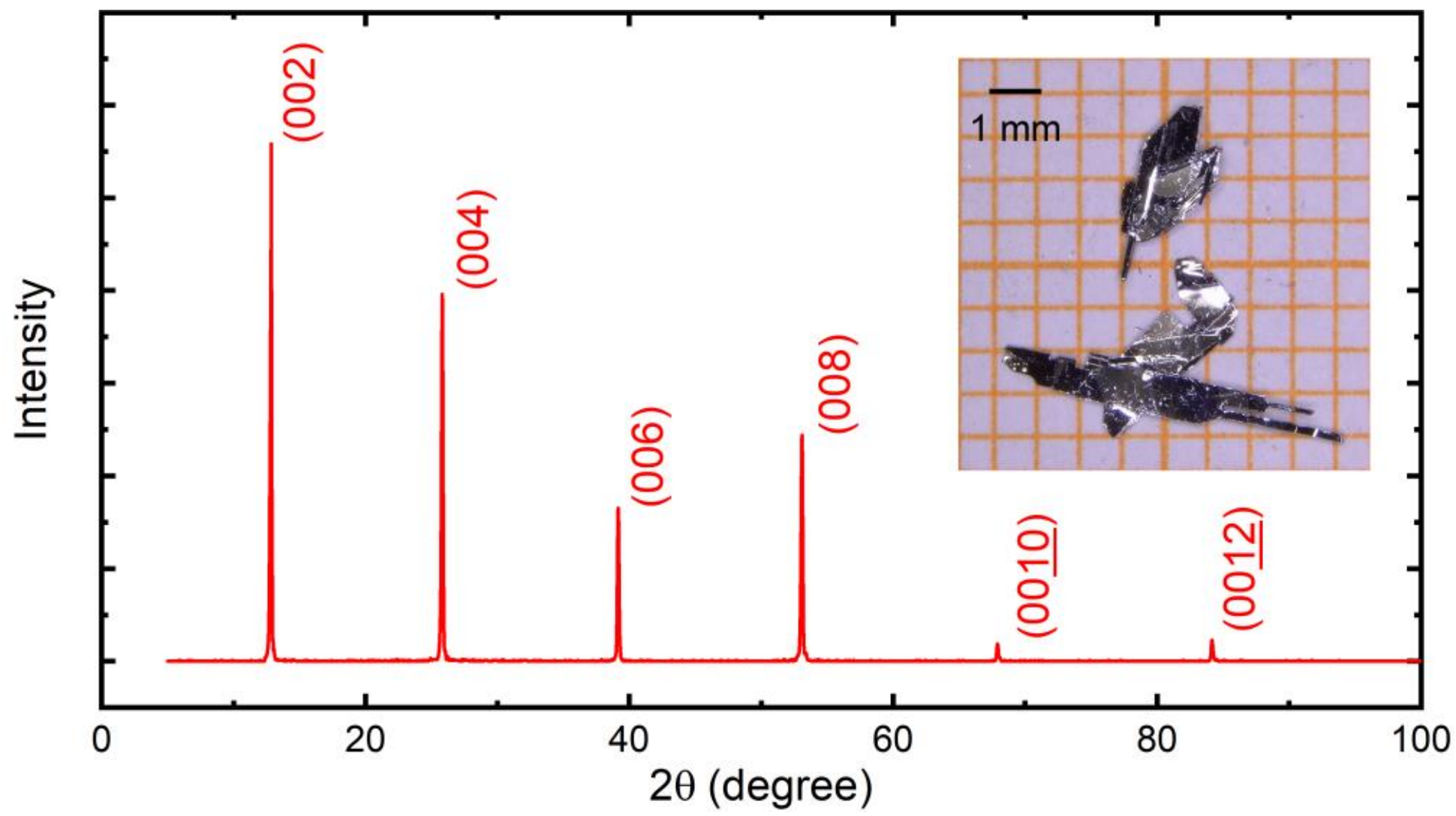

FIG. 5. XRD pattern measured from the (001) surface of $MoTe_2$ single crystal. Inset: a photograph of $MoTe_2$ single crystals.

In order to exclude the possible surface contamination or oxidation of the cleaved sample, we performed XPS measurements using the $MoTe_2$ sample from the same batch. As a control experiment, the *in situ* cleavage method in the XPS was exactly the same as the HREELS measurements. XPS was performed on the Thermo Scientific ESCALab 250Xi using 200 W monochromatic Al $K_\alpha$ radiation. The 500 μm X-ray spot was used for SAXPS analysis. The base pressure in the analysis chamber was about $3\times10^{-9}$ mbar. The hydrocarbon C1s line at 284.8 eV from adventitious carbon is used for energy referencing.

The Te 3d and Mo 3d XPS spectra are plotted in Fig. 6. The *in situ* cleaved sample show clean Te and Mo 3d peaks with no indication of oxidation or other contaminations.

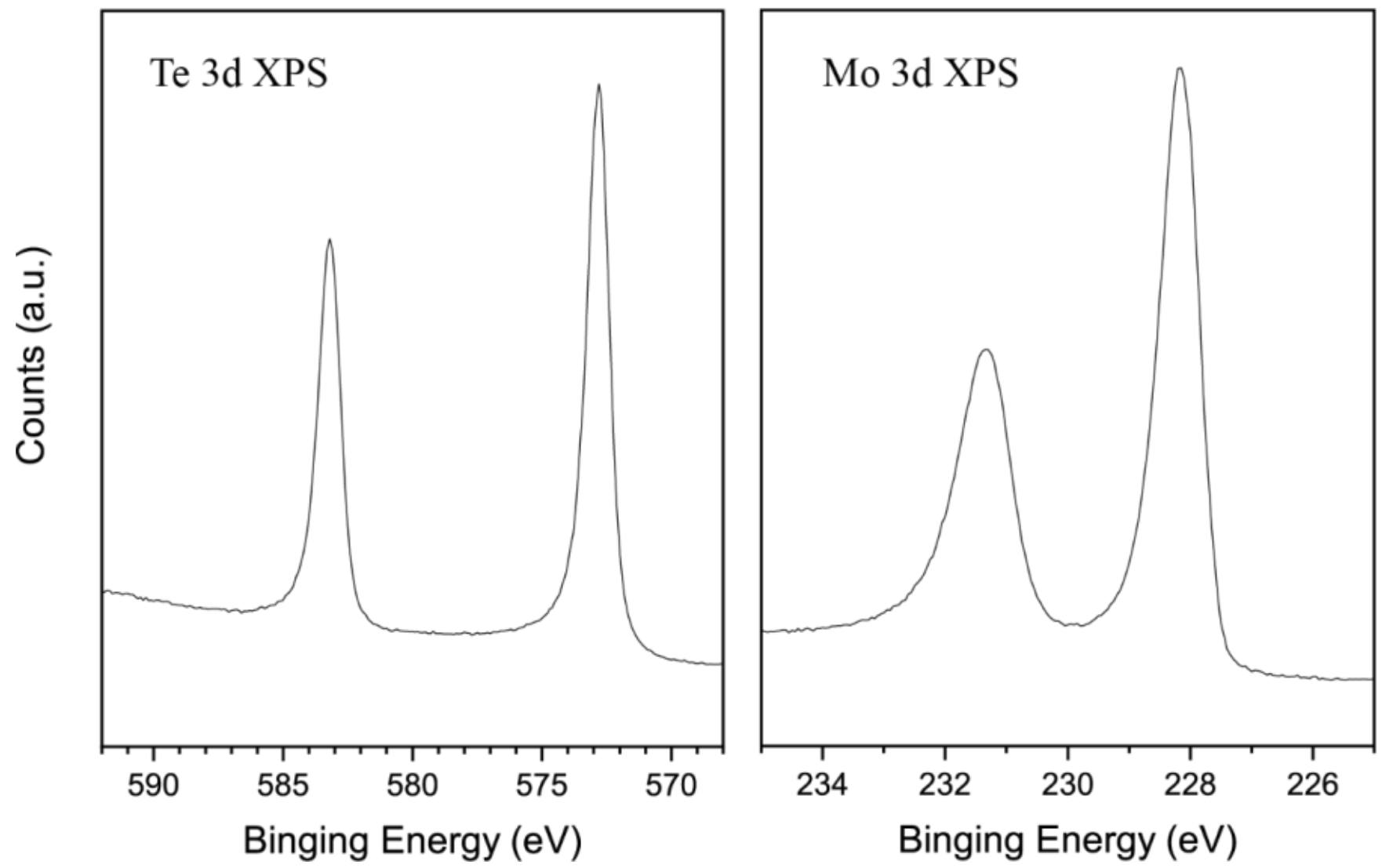

FIG. 6. XPS spectra measured from the *in situ* cleaved (001) surface of $MoTe_2$.

## Appendix B: Fitting of the HREELS spectra

The typical 2D-HREELS data set is a mapping like Fig. 2(a) or Fig. 2(b). We can extract EDCs at different momentum and fit the curves. Figure. 7 shows a typical fitting case.

Figure 7(a) shows two EDCs at $\Gamma$ point along $\Gamma - X$ direction at 294 K and 44 K, respectively. In particular, the normalized height of these energy loss peaks relative to the elastic peak is extremely low (~$10^{-4}$), which needs a rather critical analysis to obtain the intrinsic information. As $MoTe_2$ is a semimetal, there's a huge Drude background, which can be described as an analytical form

$$f(x) = A(x - x_0) + B + C(x - x_0)^{-1} + D(x - x_0)^{-2} + \ldots,$$

where the spectrum intensity $f(x)$ is a function of $x$ (energy loss) and the coefficients $x_0$, *A*, *B*, *C* and *D* to extrapolate the background. This background form was used in fitting the plasmons in graphite [54] and graphene [53]. Fitting the subtracted spectra with Lorentz lines shape can obtain the exact information of loss peaks, including energy, full width at half maximum (FWHM), height and area/intensity. Typical subtracted spectra and the fittings are illustrated in Fig. 7(b).

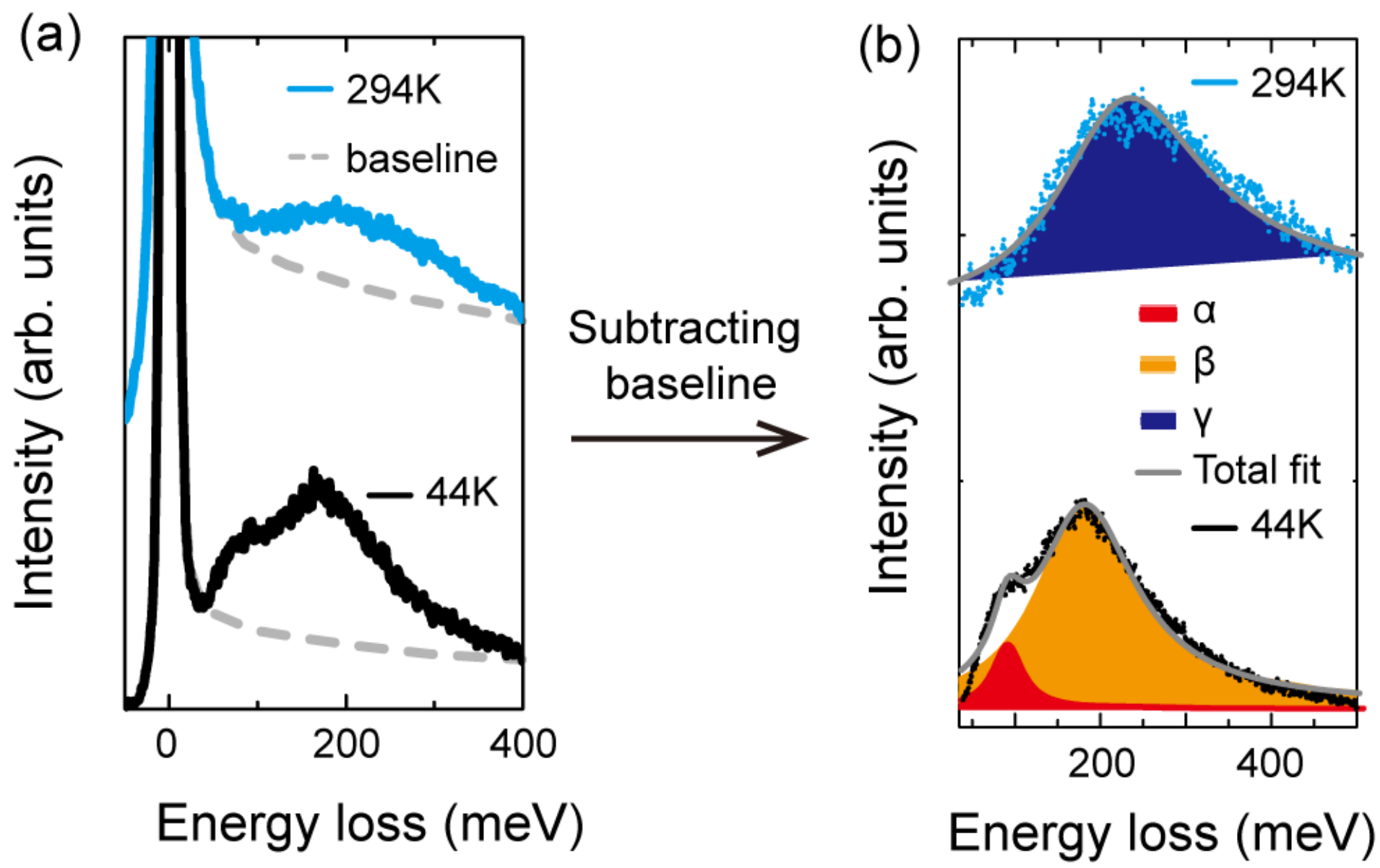

FIG. 7. (a) EDCs at Γ point with temperatures at 44 K (black line) and 294 K (light blue line). The dashed gray line is the baseline. (b) The EDCs with baselines subtracted, at temperatures 44 K and 294 K. Red, orange and dark blue solid Lorentz peaks are the α, β and γ modes, respectively. Gray solid lines are the total fits.

Different background baseline selection may generate uncertainties in the fitting results. Due to the low intensity of the energy loss peaks, we take the average value from multiple fittings and the error bars are from the uncertainty of different baselines. In Fig. 8, a typical case of the background subtraction with different baselines is demonstrated.

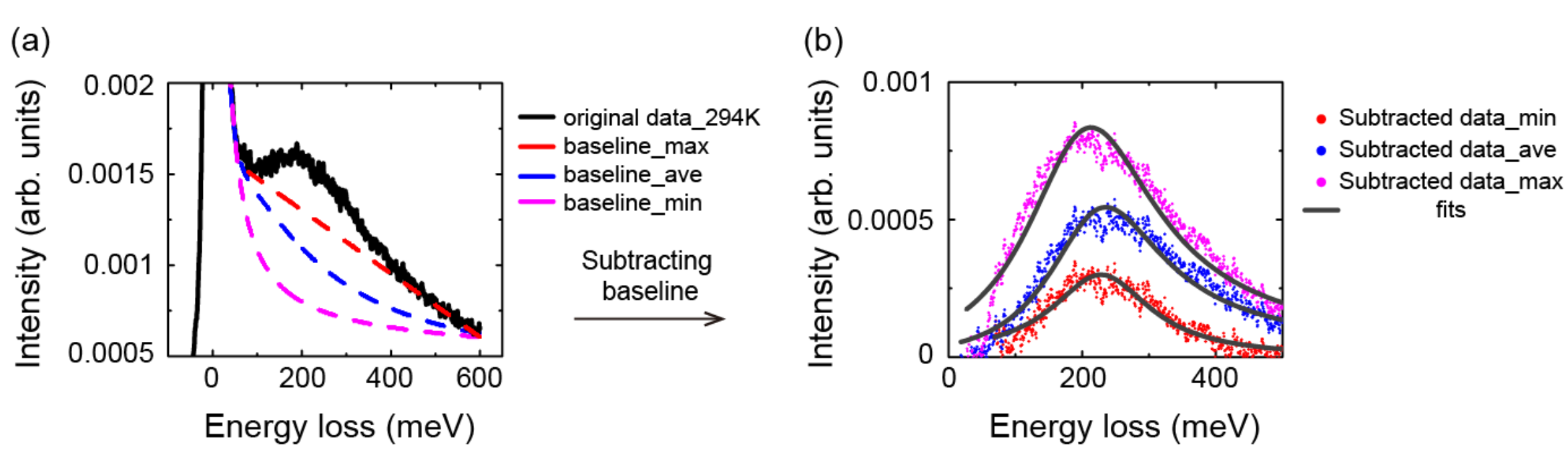

FIG. 8. A typical case of the background subtraction using different baselines.

## Appendix C: Supplementary HREELS Data

### 1. HREELS spectra along different in-plane directions

To explore if the observed plasmons are anisotropic, we performed the HREELS measurements along different in-plane directions at low temperature. The directions we chose are along $\Gamma - X$ , $\Gamma - Y$ , and about 45° between $\Gamma - X$ and $\Gamma - Y$ . In Fig. 9, the HREELS spectra along these three directions are illustrated, indicating the plasmon modes are isotropic. All other data presented in the manuscript are those collected along the $\Gamma - X$ direction, except that the data in Fig. 11 is from the $\Gamma - Y$ direction.

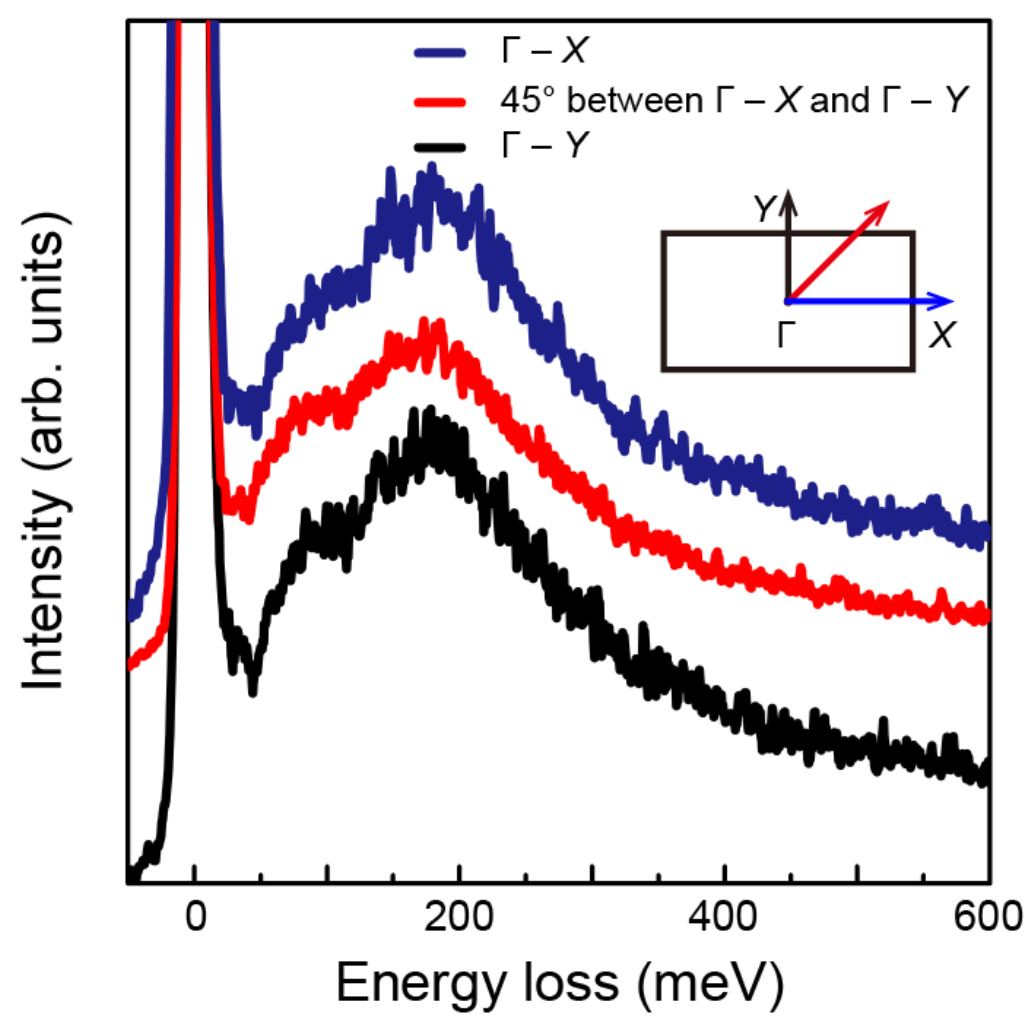

FIG. 9. Energy loss spectra around Γ point along different in-plane directions. The inset is an illustration of (001) surface BZ.

### 2. Ratio of energy between α and β in the $T_d$ phase $MoTe_2$

To further verify the nature of the measured α and β modes in the $T_d$ phase, we evaluate the energy ratio of the two modes. The results with the temperature as the x-axis are plotted in Fig. 10. The energy ratio, with an average value of $\frac{\omega_\alpha}{\omega_\beta} \sim 0.56 \pm 0.04$, is considerably smaller than the conventional ratio between the energies of bulk plasmon (bp) and its corresponding surface plasmon (sp) from the same electronic band ($\frac{\omega_{\mathrm{sp}}}{\omega_{\mathrm{bp}}} = \frac{1}{\sqrt{2}} \sim 0.71$).

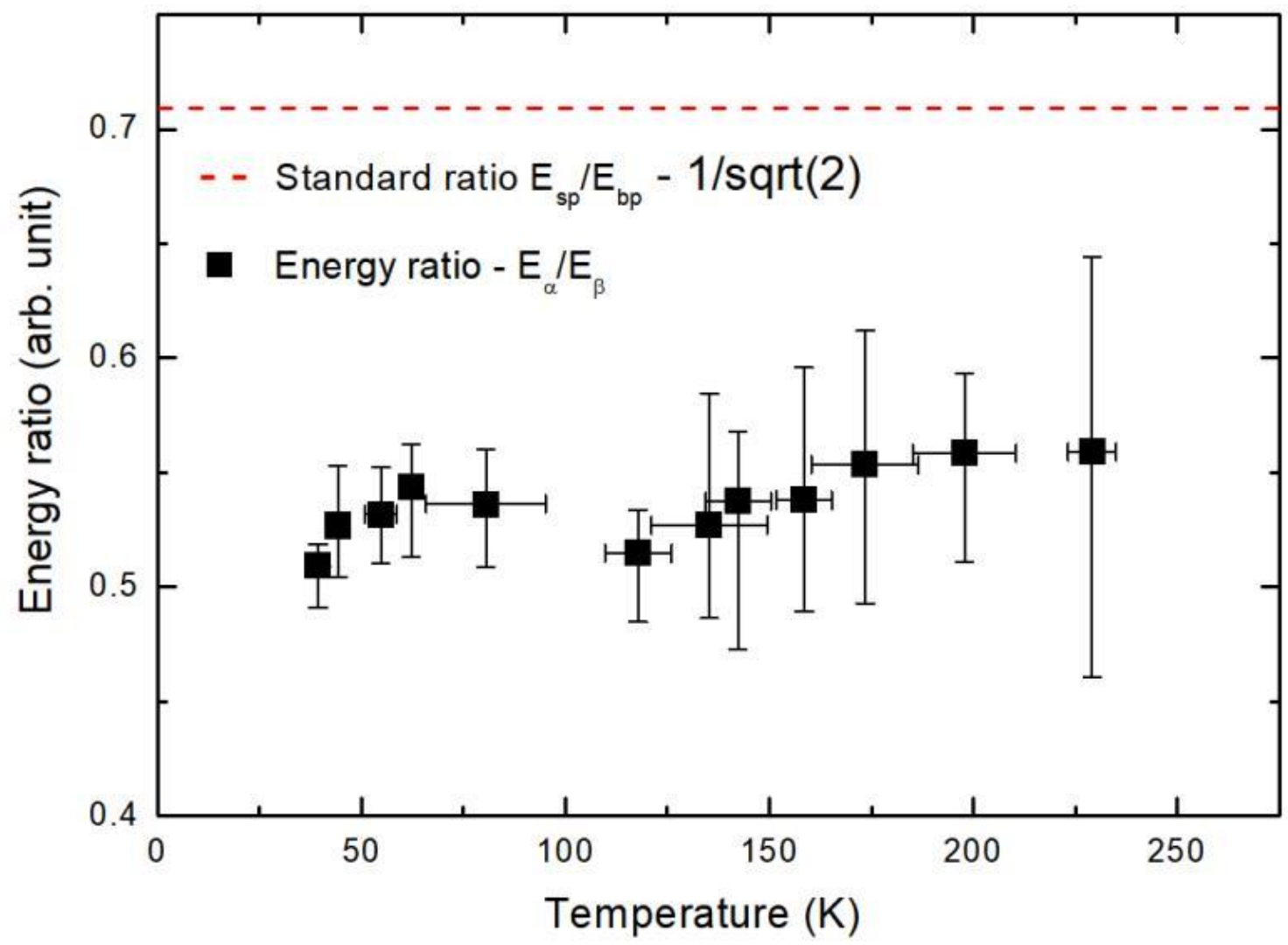

FIG. 10. The temperature-dependent energy ratio of α and β modes in the $T_d$ phase $MoTe_2$.

Moreover, the intensity ratio between the surface plasmon and the corresponding bulk plasmon should increase with the increasing incident angle (defined as the angle between the incident electron beam and the surface normal) [59], since the electrons penetrate deeper at smaller incident angle. Here, the intensity ratio of α and β modes is weakly dependent on the incident angle, which is demonstrated in Fig. 11.

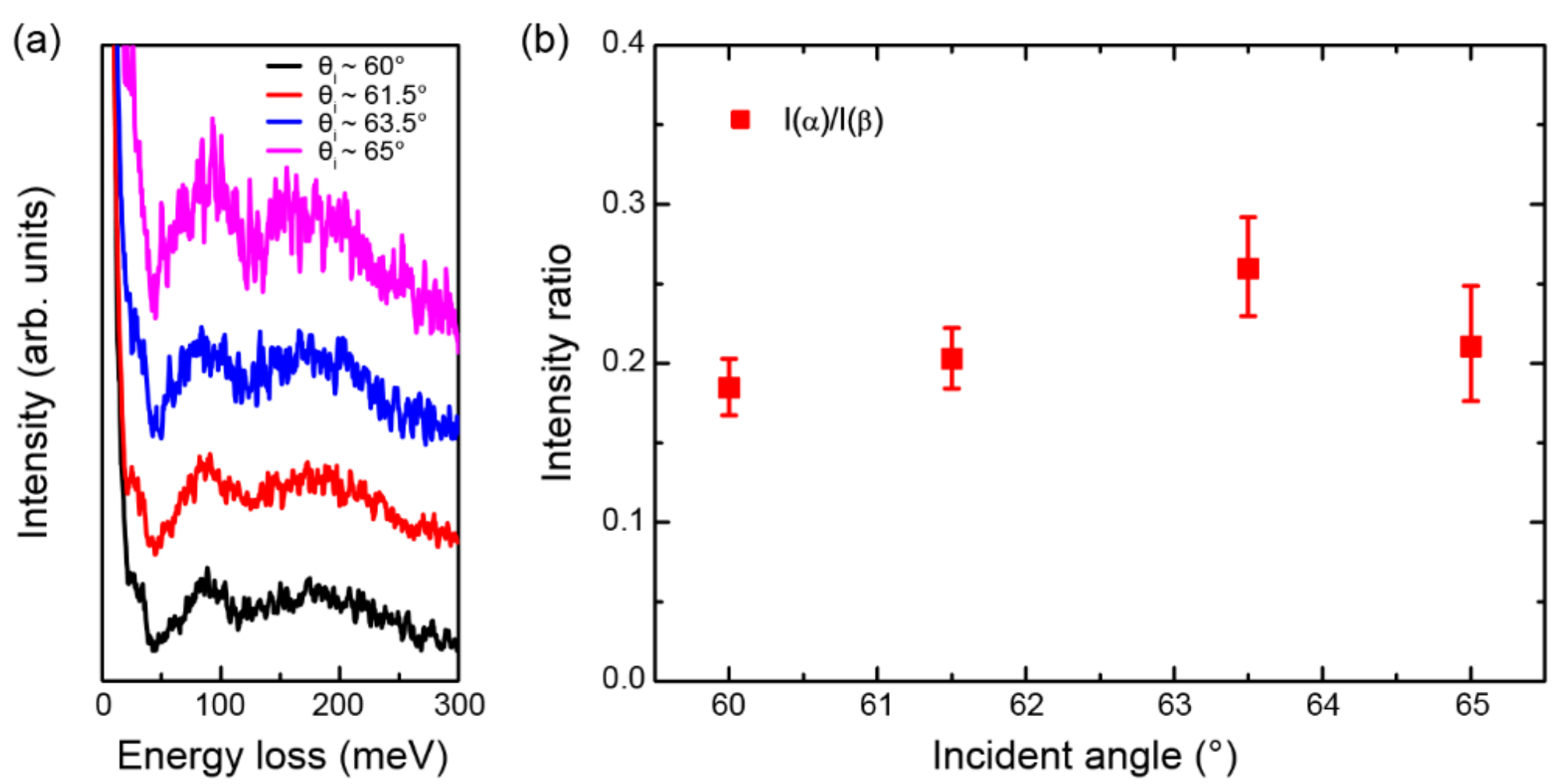

FIG. 11. HREELS measurements with different incident around Γ point at 38 K. (a) Energy loss spectra stack with different incident angles. (b) Extracted energy ratio of the α and β modes as a function of the incident angle.

These analyses suggest that β and α cannot be the conventional bulk plasmon and its corresponding surface plasmon. But it may not be rigorous to verify the nature of α and β only based on the energy ratio, since the energy ratio between the surface and bulk plasmons may become complicated when band structure effects are operative. Yet considering there is only one mode (γ) observed above 250 K, the same measurement at lower temperature observing two modes should be the result of the phase transition instead of a surface effect.

## 3. HREELS intensity check with different incident energies

We performed the HREELS measurements with the incident energy ranging from 15 eV to 110 eV at room temperature. The results are plotted in Fig. 12. The worse signal-to-noise ratio of the 50 and 80 eV data is due to the shorter data acquisition time used than the other incident energies. The peak position and intensity of the plasmon peak shows very weak dependence on the incident energy. We chose 110 eV to perform most of measurements in this study, since 110 eV is the mostly used incident energy in our facility and thus can generate the most stable beam intensity during the measurements.

The independence of the loss peaks on the incident energy can also provide auxiliary evidence that the measured modes are bulk plasmons. The mean-free path of incident electrons is strongly dependent on the incident electron energy. If the loss peak were from surface plasmon, the intensity would be strongly dependent on the incident energy.

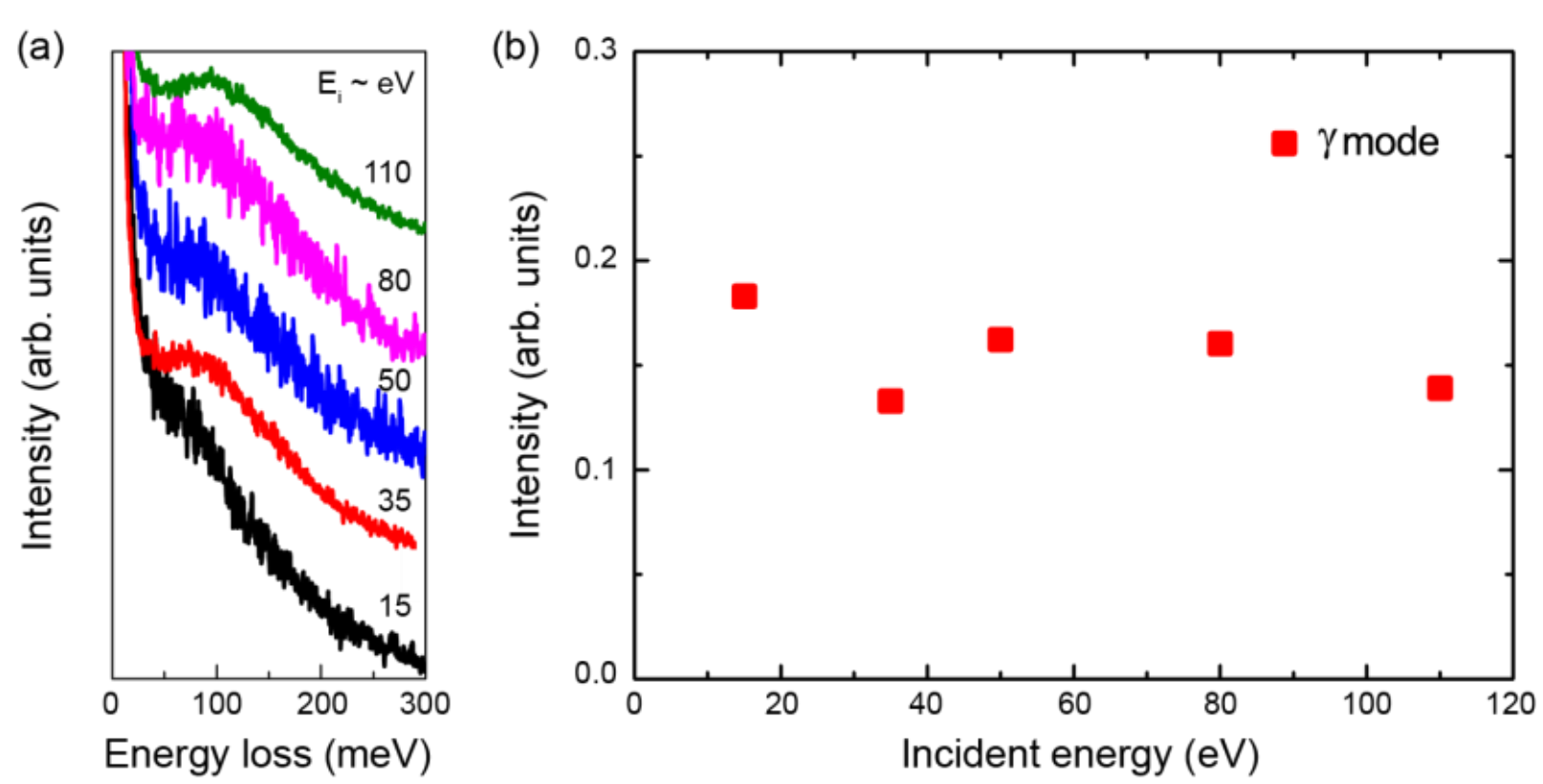

FIG. 12. HREELS measurements with different incident energy at room temperature. (a) Energy loss spectra stack with different incident energies. (b) Extracted intensity of the γ mode as a function of the incident energy.

4. Additional temperature-dependent HREELS data

In addition to the data shown in Fig. 3(a) in the main manuscript, here in Fig. 13 we also provide the temperature-dependent data with more dense temperature points, for the sake of completeness.

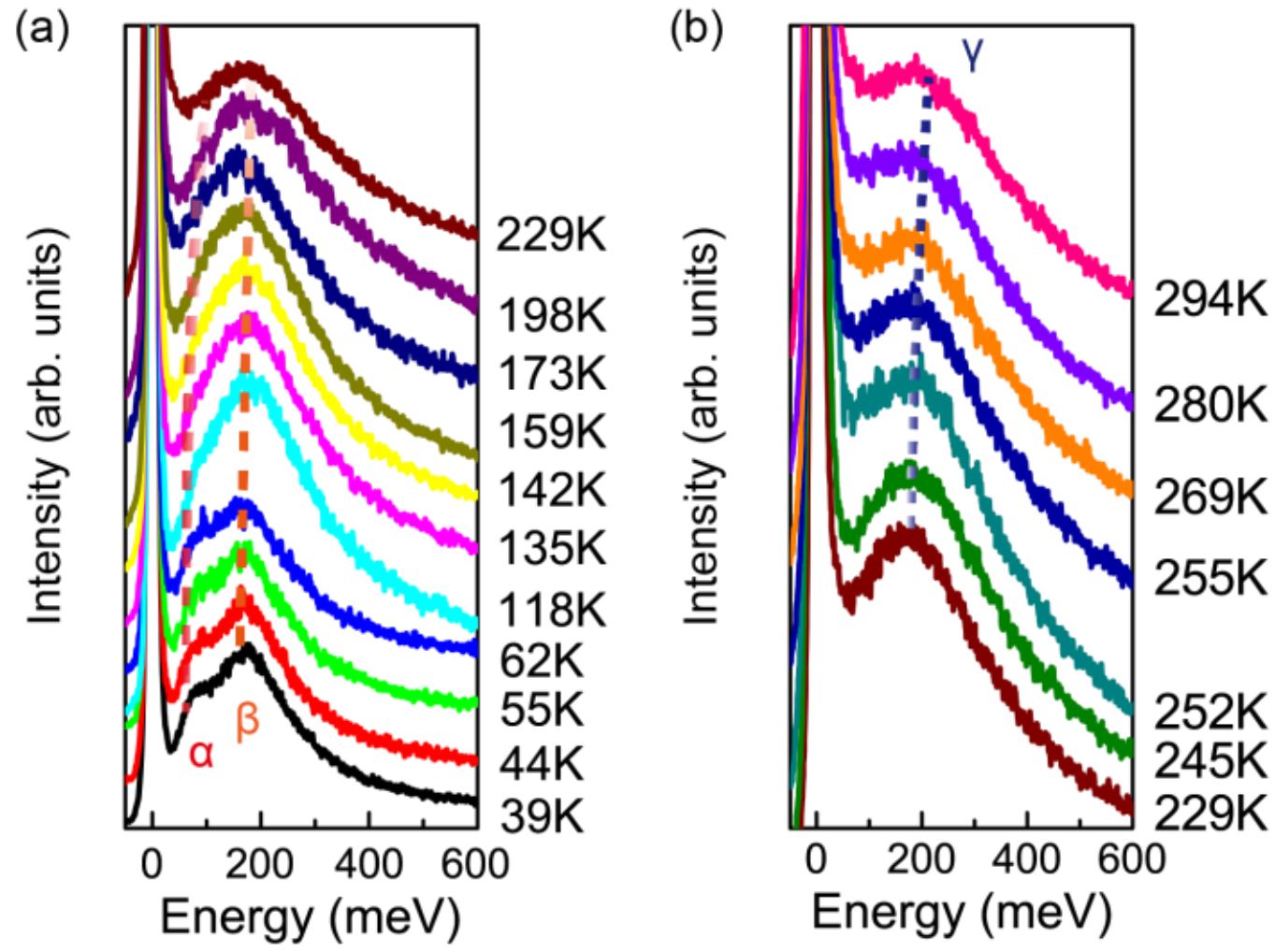

FIG. 13. Complete set of the temperature-dependent measurement. (a) EDC stacking at Γ point from 39K to 229K. (b) EDC stacking at Γ point from 229K to 294K.

## Appendix D: The Relative Strength of the Contribution from the Bands to the Plasmon Modes

The energy loss function is defined by

$$
\begin{aligned}
\varepsilon_{\text{loss}} &= -\operatorname{Im}\left[\frac{1}{\varepsilon^{\text{RPA}}(0,\omega)}\right] \\
&= -\operatorname{Im}\left[\frac{1}{\operatorname{Re}\varepsilon^{\text{RPA}}(0,\omega)+i\operatorname{Im}\varepsilon^{\text{RPA}}(0,\omega)}\right] \\
&= -\operatorname{Im}\left\{\frac{\operatorname{Re}\varepsilon^{\text{RPA}}(0,\omega)-i\operatorname{Im}\varepsilon^{\text{RPA}}(0,\omega)}{\left[\operatorname{Re}\varepsilon^{\text{RPA}}(0,\omega)\right]^2+\left[\operatorname{Im}\varepsilon^{\text{RPA}}(0,\omega)\right]^2}\right\} \\
&= \frac{\operatorname{Im}\varepsilon^{\text{RPA}}(0,\omega)}{\left[\operatorname{Re}\varepsilon^{\text{RPA}}(0,\omega)\right]^2+\left[\operatorname{Im}\varepsilon^{\text{RPA}}(0,\omega)\right]^2}
\end{aligned}
$$

where the interband contribution for $\operatorname{Im}\varepsilon^{\text{RPA}}(0,\omega)$ is

$$
\operatorname{Im}\varepsilon^{\text{RPA}}(0,\omega)=\frac{4\pi e^2}{\kappa V N_k}\sum_{m,k}\sum_{n\neq m}\left(f_{mk}-f_{nk}\right)\frac{\langle mk|\frac{\partial H}{\partial k}|nk\rangle\langle nk|\frac{\partial H}{\partial k}|mk\rangle}{(E_{nk}-E_{mk})^2}\delta\left(E_{mk}-E_{nk}+\omega\right).
$$

Based on these equations, we can identify the relative strength of the contribution of the band *m* to the plasmon mode γ (with $q$ =0, and $\omega=\omega_\gamma$) as $\sum_{n\neq m}\left(f_{mk}-f_{nk}\right)\frac{\langle mk|\frac{\partial H}{\partial k}|nk\rangle\langle nk|\frac{\partial H}{\partial k}|mk\rangle}{(E_{nk}-E_{mk})^2}\delta\left(E_{mk}-E_{nk}+\omega_\gamma\right)$ and the distribution of the relative strength of each *k*-point as $\sum_{m}\sum_{n\neq m}\left(f_{mk}-f_{nk}\right)\frac{\langle mk|\frac{\partial H}{\partial k}|nk\rangle\langle nk|\frac{\partial H}{\partial k}|mk\rangle}{(E_{nk}-E_{mk})^2}\delta\left(E_{mk}-E_{nk}+\omega_\gamma\right)$. The calculated results are shown in Fig. 4 in the main text.